\documentclass[aps,prb,twocolumn,superscriptaddress,showpacs,floatfix,amsmath,amssymb]{revtex4-1}

\usepackage{graphicx}
\usepackage{dcolumn}
\usepackage{bm}
\usepackage{amsmath} 
\usepackage[below]{placeins}
\usepackage{setspace} 

\begin{document}

This manuscript has been authored by UT-Battelle, LLC under Contract
No. DE- AC05-00OR22725 with the U.S. Department of Energy. The
United States Government retains and the publisher, by accepting the
article for publication, acknowledges that the United States
Government retains a non-exclusive, paid-up, irrevocable, world-wide
license to publish or reproduce the published form of this
manuscript, or allow others to do so, for United States Government
purposes. The Department of Energy will provide public access to
these results of federally sponsored research in accordance with the
DOE Public Access
Plan(http://energy.gov/downloads/doe-public-access-plan).

\clearpage

\title{Spin-lattice coupling mediated multiferroicity in (ND$_{4}$)$_{2}$FeCl$_{5}$$\cdot$D$_{2}$O}

\author{W. Tian}
\email{wt6@ornl.gov} \affiliation{Quantum Condensed Matter Division,
Oak Ridge National Laboratory, Oak Ridge, Tennessee 37831, USA}

\author{Huibo Cao}
\affiliation{Quantum Condensed Matter Division, Oak Ridge National
Laboratory, Oak Ridge, Tennessee 37831, USA}

\author{Jincheng Wang}
\affiliation{Quantum Condensed Matter Division, Oak Ridge National
Laboratory, Oak Ridge, Tennessee 37831, USA}

\author{Feng Ye}
\affiliation{Quantum Condensed Matter Division, Oak Ridge National
Laboratory, Oak Ridge, Tennessee 37831, USA}

\author{M. Matsuda}
\affiliation{Quantum Condensed Matter Division, Oak Ridge National
Laboratory, Oak Ridge, Tennessee 37831, USA}

\author{J.-Q. Yan}
\affiliation{Materials Science and Technology Division, Oak Ridge
National Laboratory, Oak Ridge, Tennessee 37831, USA}
\affiliation{Department of Materials Science and Engineering,
University of Tennessee, Knoxville, Tennessee 37996, USA}

\author{Yaohua Liu}
\affiliation{Quantum Condensed Matter Division, Oak Ridge National
Laboratory, Oak Ridge, Tennessee 37831, USA}

\author{V. O. Garlea}
\affiliation{Quantum Condensed Matter Division, Oak Ridge National
Laboratory, Oak Ridge, Tennessee 37831, USA}

\author{B. C. Chakoumakos}
\affiliation{Quantum Condensed Matter Division, Oak Ridge National
Laboratory, Oak Ridge, Tennessee 37831, USA}

\author{B. C. Sales}
\affiliation{Materials Science and Technology Division, Oak Ridge
National Laboratory, Oak Ridge, Tennessee 37831, USA}

\author{Randy S. Fishman}
\affiliation{Materials Science and Technology Division, Oak Ridge
National Laboratory, Oak Ridge, Tennessee 37831, USA}

\author{J. A. Fernandez-Baca}
\affiliation{Quantum Condensed Matter Division, Oak Ridge National
Laboratory, Oak Ridge, Tennessee 37831, USA}
\affiliation{Department
of Physics and Astronomy, The University of Tennessee, Knoxville,
Tennessee 37996, USA}

\date{\today}

\begin{abstract}

We report a neutron diffraction study of the multiferroic mechanism
in (ND$_{4}$)$_{2}$FeCl$_{5}$$\cdot$D$_{2}$O, a molecular compound
that exhibits magnetically induced ferroelectricity. This material
exhibits two successive magnetic transitions on cooling: a
long-range order transition to an incommensurate (IC) collinear
sinusoidal spin state at $T_{N}$=7.3~K, followed by a second
transition to an IC cycloidal spin state at $T_{FE}$=6.8~K, the
later of which is accompanied by spontaneous ferroelectric
polarization. The cycloid structure is strongly distorted by
spin-lattice coupling as evidenced by the observations of both odd
and even higher-order harmonics associated with the cycloid wave
vector, and a weak commensurate phase that coexists with the IC
phase. The appearance of the 2nd-order harmonic coincides with the
onset of the electric polarization, thereby providing unambiguous
evidence that the induced electric polarization is mediated by the
spin-lattice interaction. Our results for this system, in which the
orbital angular momentum is expected to be quenched, are remarkably
similar to those of the prototypical TbMnO$_3$, in which the
magnetoelectric effect is attributed to spin-orbit coupling.
\end{abstract}


\maketitle

``Improper multiferroics'' (also referred as type-II magnetic
multiferroics) are a unique group of materials that exhibit direct
coupling between magnetism and electric polarization
\cite{Cheong-2007,Khomskii-2009}. In these magnetically induced
multiferroics, the magnetoelectric (ME) effect is strong and the
onset of ferroelectricity arises directly from magnetic order that
breaks spatial inversion symmetry. Due to the strong ME effect,
there has been enormous interest in these materials motivated by
their potential applications in novel multifunctional devices.
However, natural single-phase magnetically driven multiferroics are
rare, only a few transition-metal oxides such as TbMnO$_3$
\cite{Kimura-TbMnO3-2003, Dagotto-2006, Xiang-2008,
Malashevich-2008,
TbMnO3-Kenzelmann-2005,Lovessey-2013,Solovyev-2011}, MnWO$_4$
\cite{MnWO4-2006}, Ni$_3$V$_2$O$_8$ \cite{Ni3V2O8-2005}, CuO
\cite{CuO-2012}, LiCuVO$_4$ \cite{Xiang-2007}, and CaCoMnO$_3$
\cite{CaCoMnO3-2008} are currently known to exhibit such effect. It
is thus of great interest to discover and investigate new materials
that will shed light on the underlying multiferroic mechanism.
Recent efforts to search for new multiferroics have been extended to
molecular compounds and metal-organic framework materials
(MOFs)\cite{Omar-2003}. In particular, several ionic salts
containing NH$_4$ have been reported to exhibit ferroelectricity
\cite{alum-ferroelectric} and the incorporation of NH$_4$ has been
used as a strategy to search for new multiferroics
\cite{Saman-2012,Xu-2011,Samantaray-2011}. In this paper, we report
a neutron diffraction study of the multiferroic mechanism in
(NH$_{4}$)$_{2}$FeCl$_{5}$$\cdot$H$_{2}$O. Using both polarized and
unpolarized neutrons, we show that the induced ferroelectricity in
(NH$_{4}$)$_{2}$FeCl$_{5}$$\cdot$H$_{2}$O is mediated via
spin-lattice coupling mechanism strikingly similar to TbMnO$_3$.

\begin{figure} [tp!]
\centering\includegraphics[width=2.8in]{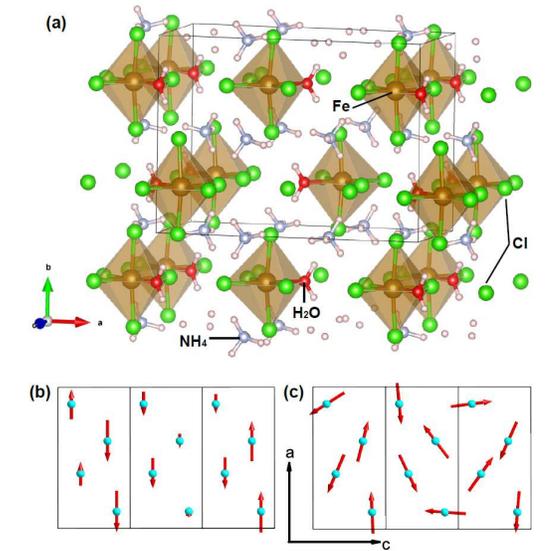}
\caption{\label{fig:magstruc}(Color online) (a) Crystal structure of
(NH$_{4}$)$_{2}$FeCl$_{5}$$\cdot$H$_{2}$O. (b) IC collinear
sinusoidal spin structure at 7~K in the paraelectric phase; (c) IC
cycloidal-spiral spin structure at 4~K in the ferroelectric phase,
as viewed along the $b$-axis (three unit cells along $c$-axis are
plotted). Only magnetic Fe ions are shown in (b) and (c) for the
purpose of clarity.}
\end{figure}

(NH$_{4}$)$_{2}$FeCl$_{5}$$\cdot$H$_{2}$O belongs to the
erythrosiderite-type compounds $A_{2}$[Fe$X_{5}$$\cdot$H$_{2}$O],
where $A$ is an alkali metal or ammonium, and $X$ is a halide ion
\cite{Carlin-1977,McElearney-1978,ref4-1995, ref3-2008}. It
crystallizes in an orthorhombic structure at room temperature (space
group \textit{Pnma}) with a crystal structure consisting of
distorted [FeCl$_{5}$$\cdot$H$_{2}$O]$^{2-}$ octahedra linked by a
network of hydrogen bonds \cite{Ackermann-2013} as illuatrated in
Fig.~\ref{fig:magstruc} (a). The orbital angular momentum of the
magnetic Fe$^{3+}$ (3d$^{5}$, high spin state) ion is expected to be
completely quenched. The magnetic interactions in these materials
are mediated via multiple superexchange pathways, such as
Fe-Cl$\cdots$Cl-Fe, Fe-O$\cdots$Cl-Fe, and Fe-O-H$\cdots$Cl-Fe
involving hydrogen bonds, suggesting the presence of magnetic
frustration in the system \cite{ref3-2008}.
(NH$_{4}$)$_{2}$FeCl$_{5}$$\cdot$H$_{2}$O \cite{Ackermann-2013} is
the only compound that exhibits spontaneous electric polarization in
the $A_{2}$[Fe$X_{5}$$\cdot$H$_{2}$O] series. In sharp contrast to
other isostructural counterparts, such as (K,
Rb)$_{2}$FeCl$_{5}$$\cdot$H$_{2}$O which undergo a single magnetic
transition adopting a collinear anferromagnetic (AFM) spin structure
with moments along the $a$-axis \cite{ref4-1995} and have no
spontaneous electric polarization,
(NH$_{4}$)$_{2}$FeCl$_{5}$$\cdot$H$_{2}$O exhibits two magnetic
transitions at $T_{N}$$\sim$7.3~K and $T_{FE}$$\sim$6.8~K,
respectively. A disorder-order transition also occurs at
$T_s$$\sim$79~K associated with the motion of the NH$_{4}$ group
\cite{Jose-2015}. Unlike other NH$_4$ contained materials
investigated so far \cite{Saman-2012,Samantaray-2011,Xu-2011} where
ferroelectricity is induced via the disorder-order transition at
high temperature decoupled with the low temperature magnetic order,
which is characteristic of the so called type-I ``proper
multiferroics'' \cite{Cheong-2007,Khomskii-2009},
(NH$_{4}$)$_{2}$FeCl$_{5}$$\cdot$H$_{2}$O remains paraelectric below
79~K and spontaneous electric polarization only appears below
$T_{FE}$. Consequently, it is crucial to know the magnetic structure
and microscopic interactions to unveil the underlying mechanism
responsible for the multiferroic properties in this compound.
However, the determination of the full magnetic structure has been
hampered due to the large amount of hydrogen atoms (40 H atoms per
unit cell) in (NH$_{4}$)$_{2}$FeCl$_{5}$$\cdot$H$_{2}$O.

\begin{figure} [tp!]
\centering\includegraphics[width=3.0in]{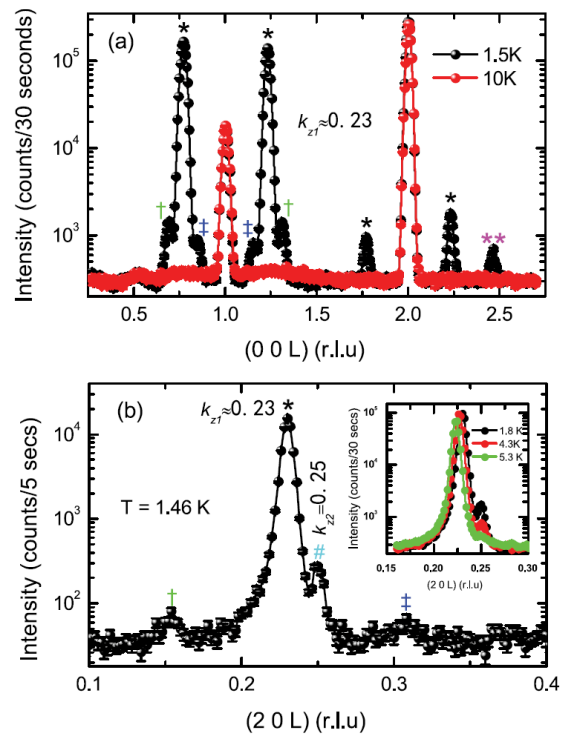}
\caption{\label{fig:scan}(Color online) L-scans along both (0 0 L)
and (2 0 L) with the intensity plotted on a logarithmic scale. (a)
(0 0 L) scan measured at 1.5~K and 10~K. The primary IC satellites
(0 0 n$\pm$$k_{z1}$), and the 2nd (0 0 n$\pm$2$k_{z1}$), 3rd (0 0
n$\pm$3$k_{z1}$), 5th order (0 0 n$\pm$5$k_{z1}$) harmonic peaks are
marked by
*(black), **(pink), $\dag$(green) and $\ddag$(blue), respectively (n is an integer). (b) (2 0 L) scan measured at 1.5~K reveals the coexistence of
IC and commensurate phases. The primary IC satellites (2 0 n$\pm$
$k_{z1}$), 3rd (2 0 n$\pm$3$k_{z1}$), 5th-order (2 0 n$\pm$
5$k_{z1}$) harmonic peaks and the commensurate peak (2 0 n$\pm$
$k_{z2}$) are marked by
*(black), $\dag$(green), $\ddag$(blue) and $\sharp$ (cyan),
respectively. Neutron data have been normalized to beam monitor
count and the error bars are statistical in nature and represent one
standard deviation.}
\end{figure}

We have grown deuterated (ND$_{4}$)$_{2}$FeCl$_{5}$$\cdot$D$_{2}$O
single crystals suitable for neutron scattering by solution method
using starting materials (DCl, FeCl$_3$, and ND$_4$Cl) similar to
that reported in Ref. \onlinecite{Ackermann-2013}. The saturated
solution was sealed and kept at 38$^{\circ}$C using a sample
environment chamber to allow slow evaporation. Large deuterated
crystals were obtained and later characterized by magnetic
susceptibility and specific heat measurements. No significant
deuteration-induced effects were observed since specific heat data
indicates no transition temperature changes compared to the
hydrogenated samples \cite{McElearney-1978,Ackermann-2013}. Elastic
neutron scattering experiments were carried out using the HB1, HB1A
triple-axis spectrometer (TAS), and HB3A Four-Circle Diffractometer
located at the High Flux Isotope Reactor (HFIR), and the Elastic
Diffuse Scattering Spectrometer (CORELLI) at the Spallation Neutron
Source (SNS) at Oak Ridge National Laboratory (ORNL). The magnetic
structures at 4~K and 7~K were determined by refining the data
collected at HB3A. HB1A and CORELLI were used to investigate the
temperature and \textit{q} dependence of the observed higher-order
harmonics associated with the cycloid order. The nature of the
higher-order harmonics were further clarified by polarized neutron
experiments using the HB1 polarized TAS. The polarization of the
incident beam and the polarization analysis of the scattered beam
were produced by Heusler crystals, and a Mezei spin flipper was used
for reversing the neutron polarization vector.

Neutron data show that (ND$_{4}$)$_{2}$FeCl$_{5}$$\cdot$D$_{2}$O
undergoes an incommensurate (IC) AFM long range order (LRO)
transition at $T_{N}$=7.3~K. The primary magnetic satellite peak can
be indexed with a propagation vector \textbf{$k_{1}$}=(0 0
$k_{z1}$), $k_{z1}$ $\approx$ 0.23 at 1.5~K, consistent with the
recent study by Rodr$\acute{i}$guez-Velamaz$\acute{a}$n et al.
\cite{Jose-2015}. To determine the magnetic structures associated
with both the paraelectric phase between $T_{FE}$$<$T$<$$T_{N}$ and
the ferroelectric phase below $T_{FE}$=6.8~K, sets of nuclear and
magnetic reflections were collected at 7~K and 4~K at HB3A (wave
length 1.546 \AA). FullProf refinement of 4~K data confirms an IC
cycloidal spiral spin structure with moments mainly confined in the
$ac$-plane (moment size $\sim$4.08~$\mu_{B}$) consistent with Ref.
\onlinecite{Jose-2015}. The magnetic structure at 7~K is found to be
an IC collinear sinusoidal spin state with moments along the
$a$-axis, with a moment size of $\sim$2.17~$\mu_{B}$. The schematic
magnetic structures at 7~K and 4~K are illustrated in
Fig.~\ref{fig:magstruc} (b) and (c), respectively. This indicates a
magnetic structure change from collinear sinusoidal to cycloidal
spiral at $T_{FE}$=6.8~K that suggests an inverse
Dzyaloshinskii-Moriya mechanism for the induced ferroelectricity as
proposed in Ref. \onlinecite{Jose-2015}. However, as we will show
below, the cycloid structure is strongly distorted via spin-lattice
coupling.

Figure 2 shows representative $L$-scans along both (0 0 L) and (2 0
L) directions. The scattering intensity is plotted on a logarithmic
scale. Fig.~\ref{fig:scan} (a) compares the $L$-scans along (0 0 L)
measured at 1.5~K and 10~K. Besides the strong, primary IC satellite
peaks, weak reflections are clearly observed at 1.5~K that can be
indexed as 2nd, 3rd and 5th-order harmonics of the wave vector
\textbf{$k_{1}$}. Furthermore, weak commensurate peaks with a
propagation vector \textbf{$k_{2}$}=(0 0 $k_{z2}$) ($k_{z2}$=0.25)
are also observed at 1.5~K. As depicted in Fig.~\ref{fig:scan} (b),
both (2 0 0.23) IC and (2 0 0.25) commensurate peaks are observed in
the $L$-scan along (2 0 L). This suggests the coexistence of
commensurate and incommensurate phases at 1.5~K. The inset
illustrates the temperature dependence of (2 0 0.23) and (2 0 0.25).
Both reflections show increased intensity with decreasing
temperature suggesting they are magnetic in origin. The observed
spin texture (both even and odd-order harmonics associated with
\textbf{$k_{1}$} and coexistence of the \textbf{$k_{2}$}
commensurate phase) indicates the cycloidal spiral spin structure is
strongly distorted since no higher-order harmonics should be
observed for an ideal cycloid spiral. Furthermore, as discussed in
Ref. \onlinecite{Kirai-1997}, odd-order harmonics and even order
harmonics are expected to be magnetic and nuclear in origin,
respectively. Therefore, the observation of the 2nd-order harmonic
provides direct evidence of a lattice modulation associated with the
cycloid order with wave vector \textbf{$k_{1}$}.

\begin{figure}
\centering\includegraphics[width=2.8in]{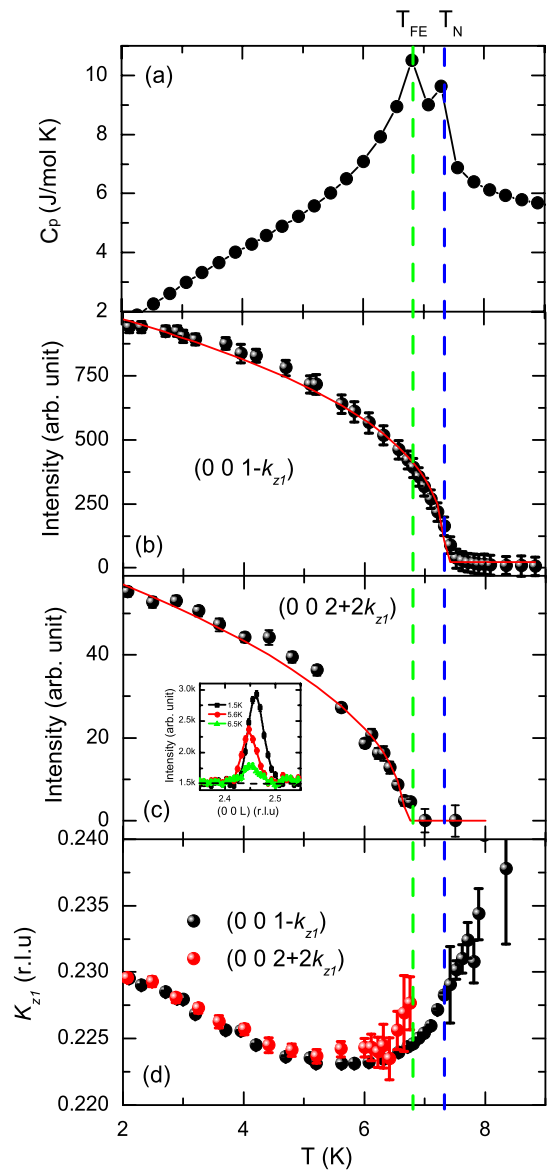}
\caption{\label{fig:op}(Color online) Temperature dependence of (a)
the specific heat, (b) integrated intensity of the (0 0 0.77)
magnetic peak, and (c) integrated intensity of the (0 0 2.46)
2nd-order harmonic peak. The inset in (c) shows $L$ scans of (0 0
2.46) at selected temperatures. The solid lines in (b) and (c) are
fits of the order-parameter data to the power law as described in
the text. (d) Temperature dependence of IC propagation vector
$k_{z1}$ determined from (0 0 1-$k_{z1}$), and (0 0 2+2$k_{z1}$).
The error bar in (d) represents statistical error from the fitting,
not the systematic error.}
\end{figure}

The magnetic transitions occur within a narrow temperature range at
$T_{N}$$\approx$7.3~K and $T_{FE}$$\approx$6.8~K. To clarify the
nature and exact transition temperature of the primary satellite and
higher-order harmonic peaks, we measure both (0 0 0.77) and (0 0
2.46) (2nd-order harmonic) as a function of temperature by
performing $L$ scans. The obtained order parameters and peak center
(indexed using $k_{z1}$) versus T are plotted in Fig.~\ref{fig:op}
in comparison with the specific heat data (Fig.~\ref{fig:op} (a)).
The integrated intensity was obtained by fitting the scan at each
temperature to a Gaussian function with a constant background. The
order parameter of (0 0 0.77) (Fig.~\ref{fig:op} (b)) shows a
transition at 7.3~K corresponding to $T_{N}$ and depicts no clear
anomaly at $T_{FE}$. The order parameter of (0 0 2.46)
(Fig.~\ref{fig:op} (c)) reveals that the appearance of the 2nd-order
harmonic coincides with the onset of electric polarization. Fitting
the order parameters to a power-law
\textit{I(T)=I$_0$[(T$_{N}$-T)/T$_{N}$)]$^{2\beta}$} yield
$T_{N}$$\approx$7.34$\pm$0.02~K and $\beta$$\approx$0.194$\pm$0.015
for (0 0 0.77), and $T_{FE}$$\approx$6.68$\pm$0.03~K and
$\beta$$\approx$0.245$\pm$0.002 for (0 0 2.46), respectively. The
fitting results are plotted in Fig.~\ref{fig:op} (b) and (c) as red
solid curves. The fitting was performed over the temperature ranges
2~K$<T<T_{N}$ and 2~K$<T<T_{FE}$, respectively. These critical
exponent $\beta$ values are much smaller than the theoretical value
of 0.36 predicted for a three-dimensional Heisenberg antiferromagnet
(de Jongh and Miedema 1974). This suggests a lower magnetic
dimensionality of (ND$_{4}$)$_{2}$FeCl$_{5}$$\cdot$D$_{2}$O as
proposed in Ref. \onlinecite{McElearney-1978}. Temperature
dependence of the IC propagation vector $k_{z1}$, determined by
fitting the $L$-scans of (0 0 0.77) and (0 0 2.46), is plotted in
Fig.~\ref{fig:op} (d). Good agreement is obtained between
2~K$<$$T$$<$6~K, the discrepancy above 6~K can be attributed to the
significant broadening of both reflections near $T_{FE}$ and
$T_{N}$.  The wave vector (0 0 $k_{z1}$) continues to vary with
decreasing temperature throughout the ferroelectric phase and it
shows no evidence of an incommensurate-commensurate (IC-C) lock-in
transition down to 1.5~K.

\begin{figure} [tp!]
\centering\includegraphics[width=2.8in]{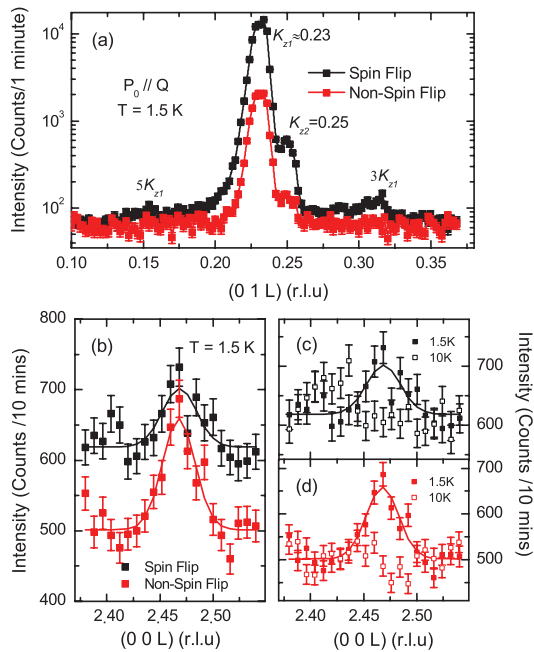}
\caption{\label{fig:hb1}(Color online) Polarized neutron data
measured in the horizontal field configuration, P$_0$ $\parallel$ Q.
(a) L-scan along [0 1 L] measured at 1.5~K comparing the spin-flip
(SF, -+) and non-spin-flip (NSF, ++) scatterings with the intensity
plotted on a logarithmic scale. (b) Comparison of the 1.5~K SF and
NSF data suggests the hybrid nature of the (0 0 2.46) 2nd-order
harmonic peak. Comparison of 1.5~K and 10~K data of (0 0 2.46) from
(c) SF channel and (d) NSF channel, respectively.}
\end{figure}

To clarify the origin of the higher-order harmonics and $k_{2}$-type
commensurate peaks, a polarized neutron diffraction experiment was
carried out using the HB1 polarized triple-axis spectrometer.
Fig.~\ref{fig:hb1} plots the polarized data analyzed using the
method discussed in Ref.
\onlinecite{Moon-1969,Huang-1992,Lynn-1993}. The magnetic origin of
the commensurate peak and the odd-order harmonics associated with
the cycloid order is verified by the stronger scattering observed in
the spin-flip (SF, -+) channel in Fig.~\ref{fig:hb1} (a). The
remaining scattering in the non-spin-flip (NSF, ++) channel can be
attributed to finite instrumental flipping ratio which is estimated
to be $\sim$1/10 by comparing the integrated intensity of SF and NSF
data of the (0 0 2) nuclear peak. Fig.~\ref{fig:hb1} (b) compares
the SF and NSF data of (0 0 2.46) measured at 1.5~K. Stronger
scattering intensity is detected in the NSF channel indicating the
2nd-order harmonic is dominated by nuclear scattering contribution.
Weak scattering is also observed in the SF channel. The integrated
intensity ratio between NSF and SF is $\sim$1.8, much smaller than
the ratio of $\sim$10 obtained for the nearby (0 0 2) nuclear peak,
suggesting a hybrid nature displaying both nuclear and magnetic
characters. Fig.~\ref{fig:hb1} (c) and (d) compares 1.5~K and 10~K
data measured with SF and NSF configurations, respectively. In both
cases, the (0 0 2.46) peak vanishes at 10~K in good agreement with
the order parameter results measured with unpolarized neutrons.

The spin texture (in the form of higher-order harmoincs) observed in
(ND$_{4}$)$_{2}$FeCl$_{5}$$\cdot$D$_{2}$O is strikingly similar to
that of TbMnO$_3$, a well studied type-II multiferroic system
\cite{Kimura-TbMnO3-2003, Dagotto-2006, Xiang-2008,
Malashevich-2008,
TbMnO3-Kenzelmann-2005,Lovessey-2013,Solovyev-2011}. Three
transitions are observed in TbMnO$_3$ at 41~K, 27~K, and 7~K
\cite{Kimura-TbMnO3-2003}. The first one at 41~K is associated with
the IC magnetic LRO transition of Mn$^{3+}$, and the third one at
7~K is due to the ordering of Tb$^{3+}$ moments, typical for
rare-earth compounds. The second transition at 27~K is particularly
interesting and is accompanied by a dielectric anomaly and electric
polarization. It has been clarified by a neutron diffraction study
\cite{TbMnO3-Kenzelmann-2005} that this 27~K transition is
associated with a magnetic structure change from an IC sinusoidally
modulated collinear magnetic structure to an IC noncollinear spiral.
Strong odd-order harmonics are observed indicating the spiral spin
configuration is strongly distorted. In particular, the observation
of a 2nd-order harmonic associated with chiral wave vector
\cite{Kimura-TbMnO3-2003} ignited a debate of the FE mechanism
between the ``pure electronic'' model by Katsura, Nagaosa, and
Balatsky \cite{Katsura-2005} that suggests the electric polarization
can be induced without the involvement of lattice degrees of
freedom, and the ``ion displacement'' model that suggests a lattice
distortion via spin-orbit coupling is essential for the FE
polarization \cite{Dagotto-2006,Xiang-2008,
Malashevich-2008,Solovyev-2011}. Resonant $x$-ray diffraction study
\cite{Lovessey-2013} reveals the melting of the chiral order
associated with the onset of the electric polarization at 27~K, in
which the 2nd-order harmonic is unambiguously detected with an order
parameter behaving similarly to what we observed in
(ND$_{4}$)$_{2}$FeCl$_{5}$$\cdot$D$_{2}$O. The intensity of the
2nd-order harmonic increases sharply at 27~K coinciding with the
onset of the ferroelectricity. These experimental results combined
with theory \cite{Dagotto-2006,Xiang-2008,
Malashevich-2008,Solovyev-2011} have classified TbMnO$_3$ as a
prototypical material in which a cycloidal-spin structure generates
an electric polarization via the spin-orbit interaction.

Comparing our results to TbMnO$_3$, it is quite remarkable that
almost identical behavior (spin texture) is observed in a molecular
compound like (ND$_{4}$)$_{2}$FeCl$_{5}$$\cdot$D$_{2}$O in which the
orbital angular momentum of Fe$^{3+}$ is expected to be completely
quenched. Both even and odd-order harmonics are observed. Similarly,
the magnetic structure of (ND$_{4}$)$_{2}$FeCl$_{5}$$\cdot$D$_{2}$O
changes from IC collinear sinusoidal to IC cycloidal-spiral at
$T_{FE}$ without an IC-C lock-in. The material is paraelectric
between $T_{FE}<T<T_N$ and ferroelectricity only sets in below
$T_{FE}$, where the 2nd-order harmonic appears indicating lattice
distortion resulting from spin-lattice coupling. Hence, our neutron
diffraction study provides unambiguous evidence that spin-lattice
coupling plays a key role in its FE mechanism. Moreover, the
observed coexistence of incommensurate and commensurate phases adds
additional complexity in this material compared with TbMnO$_3$. A
preliminary inelastic neutron scattering study suggests they are
both energetically favorable and the competition between these two
phases is responsible for the rich magnetic field versus temperature
phase diagram of (ND$_{4}$)$_{2}$FeCl$_{5}$$\cdot$D$_{2}$O
\cite{wei-unpublished}.

In summary, our neutron diffraction study on
(ND$_{4}$)$_{2}$FeCl$_{5}$$\cdot$D$_{2}$O reveals that the
appearance of ferroelectricity in this molecular magnet is mediated
by spin-lattice coupling. We find remarkable similarities with
TbMnO$_3$. Extensive experimental
\cite{Kimura-TbMnO3-2003,Kimura-TbMnO3-2003,Lovessey-2013} and
theoretical studies \cite{Dagotto-2006,Xiang-2008,
Malashevich-2008,Solovyev-2011} indicate that the subtle distorted
spin texture observed in TbMnO$_3$ plays a critical role in the FE
mechanism for this prototypical system. Our study suggests
(ND$_{4}$)$_{2}$FeCl$_{5}$$\cdot$D$_{2}$O is a rare molecular
analogue to TbMnO$_3$. Although the incorporation of NH$_4$ has been
used as a strategy to search for new multiferroics, however, to our
knowledge only type-I multiferroics have been discovered through
this approach so far \cite{Saman-2012,Xu-2011,Samantaray-2011}. The
spin-lattice coupling mediated FE mechanism revealed in this system
certainly opens an avenue for searching new multiferroics and calls
for more thorough theoretical investigations.

\section{Acknowledgement}

The research work at ORNL's High Flux Isotope Reactor and Spallation
Neutron Source was sponsored by the Scientific User Facilities
Division, Office of Basic Energy Sciences, US Department of Energy.
JQY, BCS and RSF were supported by the Department of Energy, Office
of Science, Basic Energy Sciences, Materials Sciences and
Engineering Division.

\end{document}